\documentclass[pra,aps,showpacs,groupedaddress,superscriptaddress,twocolumn,toc=flat]{revtex4-1}

\usepackage[colorlinks=true,linkcolor=blue,urlcolor=blue,citecolor=blue]{hyperref}

\usepackage[utf8x]{inputenc}
\usepackage{color}
\usepackage{bbm} 

\usepackage{amsfonts,amsmath,amssymb,stmaryrd}

\usepackage{graphicx}
\usepackage{subfigure}  
\usepackage{bbm} 
\usepackage{hyperref}
\usepackage{epsfig}
\usepackage{mathrsfs}
\usepackage{verbatim}
\usepackage{centernot}
\usepackage{ulem}
\usepackage{nicefrac}

\renewcommand{\l}{\left(}
\renewcommand{\r}{\right)}

\newcommand{\bra}[1]{\langle#1|}
\newcommand{\ket}[1]{|#1\rangle}

\renewcommand{\ij}{{\langle \vec{i}, \vec{j} \rangle}}

\renewcommand{\H}{\hat{\mathcal{H}}}

\renewcommand{\c}{\hat{c}}

\newcommand{\cd}{\hat{c}^\dagger}

\newcommand{\hc}{\text{h.c.}}

\usepackage{array}

\usepackage{cancel,ifthen}
\newcommand{\cmnt}[2][NoInPuT]{\ifthenelse{\equal{#1}{NoInPuT}}{}{{\color{red}\sout{#1}}} {\color{blue} #2}}

\usepackage{bm}	
\renewcommand{\vec}[1]{\bm{#1}}

\begin{document}
\normalem	

\title{Spectroscopy of Hubbard-Mott excitons and their ro-vibrational excitations}

\author{A. Bohrdt}
\email[Corresponding author email: ]{annabelle.bohrdt@ur.de}
\affiliation{Institut für Theoretische Physik, Universität Regensburg, D-93035 Regensburg, Germany}
\affiliation{Munich Center for Quantum Science and Technology (MCQST), Schellingstr. 4, D-80799 M\"unchen, Germany}

\author{E. Demler}
\affiliation{Institut f\"{u}r Theoretische Physik, ETH Zurich, 8093 Zurich, Switzerland}

\author{F. Grusdt}
\affiliation{Department of Physics and Arnold Sommerfeld Center for Theoretical Physics (ASC), Ludwig-Maximilians-Universit\"at M\"unchen, Theresienstr. 37, M\"unchen D-80333, Germany}
\affiliation{Munich Center for Quantum Science and Technology (MCQST), Schellingstr. 4, D-80799 M\"unchen, Germany}

\date{\today}

\begin{abstract}
Hubbard excitons are bound states of doublons and holes that can be experimentally probed both in real materials, such as cuprates, and in cold atom quantum simulators. 
Here we compare properties of a Hubbard exciton to those of a pair of distinguishable dopants in the $t-J$ model and show how  insights into pair properties can be obtained through excitonic spectra. In particular, we perform large-scale numerical simulations of spectral functions and optical conductivities and obtain remarkable agreement between Hubbard excitons and pairs of distinguishable dopants. The latter can be decomposed into symmetric (bosonic) and anti-symmetric (fermionic) sectors of indistinguishable dopants, thus enabling a detailed understanding of different features observed in the excitonic spectra through comparison with a semi-analytical geometric string theory approach. We further compare theoretically computed exciton spectra in a single band Fermi-Hubbard model to resonant inelastic X-ray scattering (RIXS) studies of the parent insulating cuprate materials. We find remarkable agreement between the two spectra in both energy and momentum dependence. Our analysis suggests that multiple long-lived ro-vibrational exciton resonances have been observed in RIXS spectra. Experimentally, these features are known to persist up to optimal doping. The comparison we provide between semi-analytical theory, large-scale numerics, and experimental data thus provides an explanation of the RIXS measurements and provides new insight into the nature of pairing in cuprates.
\end{abstract}

\maketitle

\section{Introduction}
An exciton is a bound state of an electron and a hole and is ubiquitous in systems of interacting electrons, ranging from Wannier excitons \cite{Wannier1937} with comparably large spatial extend to very tightly bound Frenkel excitons \cite{Frenkel1931}, which are common in organic semiconductors. Conventional excitons are held together by the Coulomb interaction between their two constituents.  

In strongly correlated electronic systems, the exciton binding mechanism can be more complicated, and the simple Coulomb picture, which enables a description akin to a hydrogen atom, needs to be extended. 
In particular, in two-dimensional antiferromagnetic (AFM) Mott insulators, apart from the Coulomb interaction, the underlying antiferromagnetic spin correlations can lead to additional attraction of electrons and holes. The existence and properties of excitons in these systems thus potentially enables insights into the complex interplay between spin and charge degrees of freedom, as well as the role of Coulomb forces. Unraveling the structure of excitons in such strongly correlated systems is tightly connected to the quest for understanding the origin and mechanism of pairing in uncoventional superconductors, such as the cuprate materials \cite{Keimer2015}.  

Here we perform large-scale numerical simulations of excitonic spectra in the two-dimensional Fermi-Hubbard model on a four-leg cylinder using time-dependent matrix product states (t-MPS). We compare the Fermi-Hubbard excitonic spectra to the spectra for two dopants in the $t-J$ model, and establish a close relationship between the two. Existing as well as future measurements probing excitonic properties can thus shed light on the internal structure of pairs in Fermi-Hubbard type models. We compare our numerical results to solid state measurements, and discuss potential probes using cold atom quantum simulators \cite{Bohrdt2021_review}. 

The mechanism by which pairs of charge carriers form in strongly repulsive microscopic models, in particular in the two-dimensional Fermi-Hubbard model and its descendant, the $t-J$ model, has been extensively studied since the discovery of unconventional superconductivity in the cuprate compounds. Key questions in the field concern the pairing symmetry and mechanism,  the resulting binding energies, and the competition as well as interplay between pair formation and other orders, such as charge density waves \cite{Scalapino2012,Fradkin2015,Qin2019}. 
Experimentally, the pairing symmetry in the cuprates was determined to be $d_{x^2-y^2}$, corresponding to the $C_4$ angular momentum $m_4=2$, fairly early on through the temperature dependence of the London penetration depth \cite{Hardy1993} and the nuclear relaxation rate \cite{Hammel1989}, the superconducting gap anisotropy observed in angle-resolved photoemission spectroscopy (ARPES) \cite{Shen1993,Shen1993b}, the field dependence of the specific heat \cite{Moler1997}, and experiments involving two separate Josephson junctions in $x$- and $y$-direction \cite{Wollman1993}; see also \cite{Scalapino1995}. Recent analytical \cite{Grusdt2022} and numerical \cite{Bohrdt2023} work has established an extension of the single-particle spectral function, as accessed through ARPES, to pairs of dopants as a valuable tool to probe the structure of bound states, including their dispersion. Crucially, different angular momenta $m_4$ can be imparted on the system, thus enabling direct access to pairs with different pairing symmetries. Experimentally, these pair spectra could in principle be accessed through coincidence ARPES, as proposed in ~\cite{Berakdar1998,Su2020,Mahmood2022}. Here we propose studies of excitonic excitations as a powerful alternative to access properties of pairs of charge carriers.

It has been pointed out early on \cite{Clarke1993,Wrobel2002} that the strong magnetic interaction in weakly doped antiferromagnetic Mott insulators can provide a mechanism of the exciton binding energy and, as a consequence, that the analysis of Hubbard excitons might yield insights into the nature of hole pairing in these systems. 
Here we take this analysis one step further and establish a quantitative correspondence between excitonic properties and properties of pairs of indistinguishable charge carriers. In particular, we numerically calculate excitonic spectra $A_{m_4}^{(FH)}(k,\omega)$ with momentum and frequency resolution for different values of the $C_4$ angular momentum $m_4$ in the single-band Fermi-Hubbard model. 

\begin{figure}
\centering
\includegraphics[width=0.95\linewidth]{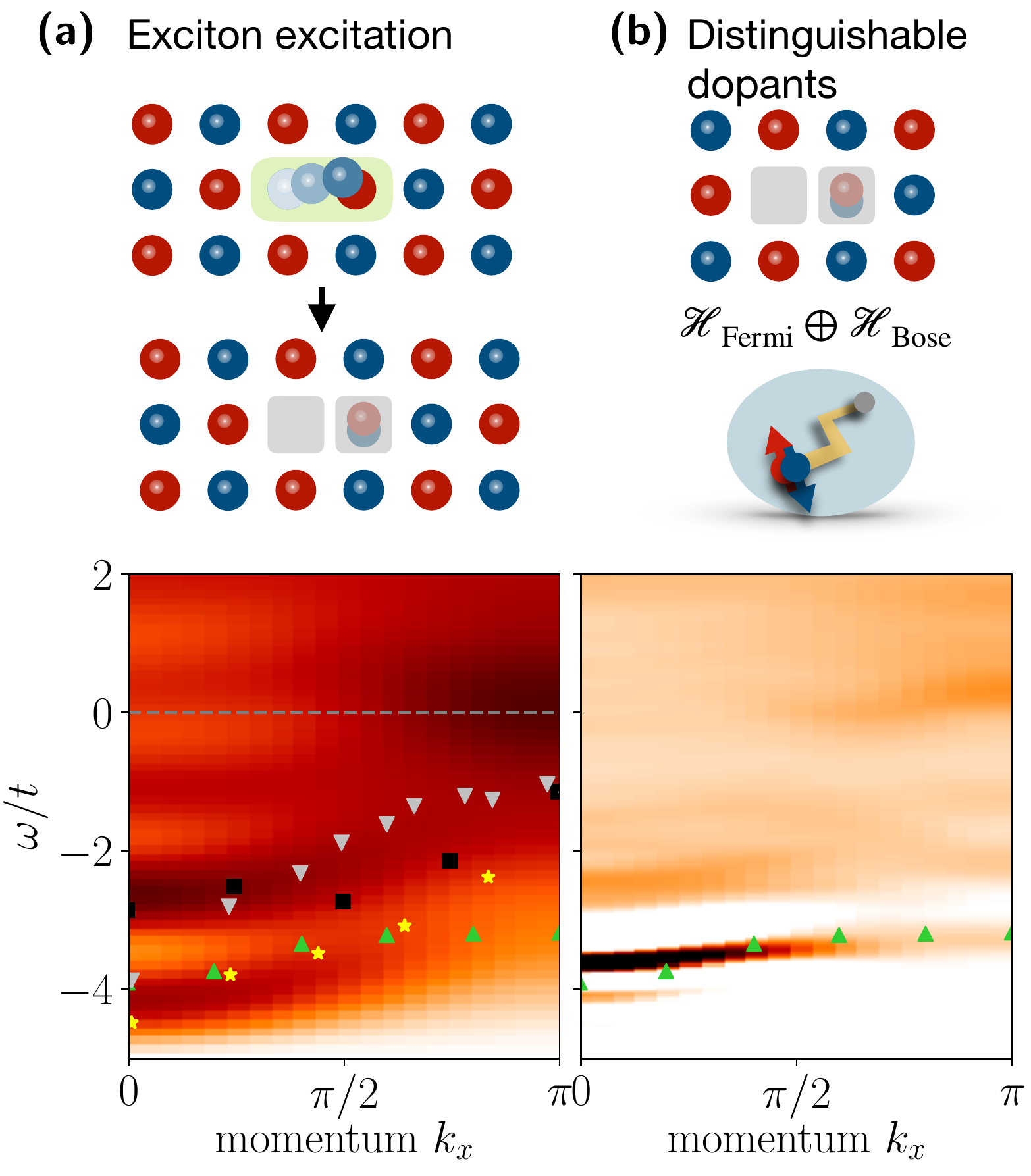} 
\caption{
\textbf{Exciton spectra} for a $d$-wave ($m_4=2$) excitation with momentum resolution in $x$-direction at $k_y=0$; (a) for the Fermi-Hubbard model at $U/t=12$, and (b) for two distinguishable dopants in the $t-J$ model at $t/J=3$ (b).  Symbols show the peak positions as extracted from 
RIXS measurements on $\text{Ca}_2\text{CuO}_2\text{Cl}_2$ \cite{Hasan2000} (squares), $\text{La}_2\text{CuO}_4$ (up triangles from \cite{Ellis2008}, stars from \cite{Collart2006}, down triangles from \cite{Kim2002}) assuming $t=350\text{meV}$ and using an energy offset of $\Delta E=10.2 t$, consistent with ARPES results \cite{Wells1995}. 
}
\label{fig:spectrum_numericsRIXS}
\end{figure}

In contrast to pairs of holes, the constituents of an exciton in an AFM Mott insulator -- a doublon and a hole -- are distinguishable. We therefore compare the spectra of doublon hole pairs at large Hubbard interactions to the spectral function of a pair of distinguishable dopants in the $t-J$ model,  $A_{m_4}^{(d-t-J)}(k,\omega)$, and find good agreement for peak positions and their dispersion, see Fig.~\ref{fig:spectrum_numericsRIXS}. The numerically obtained dispersion we find is furthermore consistent with resonant inelastic x-ray scattering (RIXS) measurements on $\text{La}_2\text{CuO}_4$ \cite{Ellis2008}. 
Based on our understanding of pairs of dopants in $t-J$-type models \cite{Shraiman1988,Grusdt2022} in terms of the geometric string theory \cite{Grusdt2019}, we thus establish a string-like internal structure of Hubbard excitons. This view is further supported by the observation of string-like ro-vibrational resonances in exciton spectra \cite{Ellis2008,Lu2005,Taranto2012}

\begin{figure}
\centering
\includegraphics[width=0.99\linewidth]{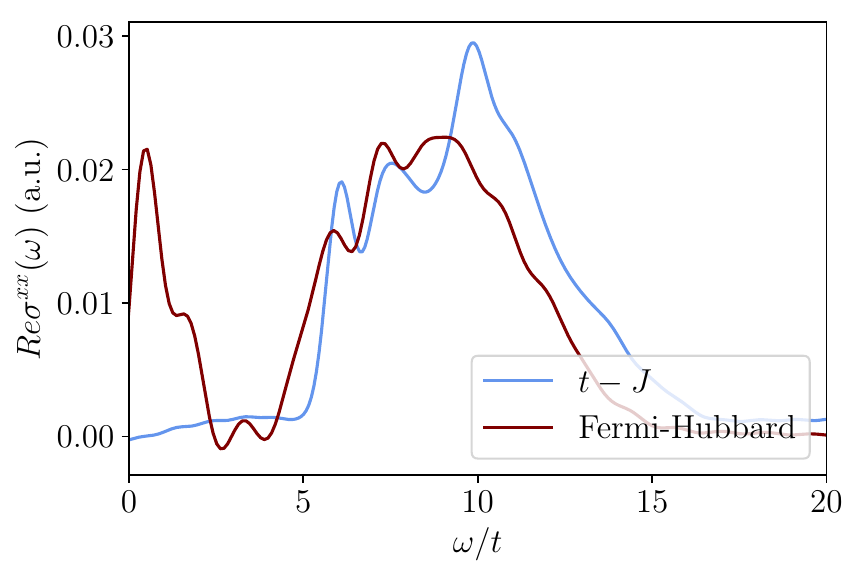}
\caption{
\textbf{Optical conductivity} in the Fermi-Hubbard model (red) for $U/t=8$ compared to the corresponding response function for a pair of two distinguishable holes in the $t-J$ model (blue) with $t/J=2$.  
}
\label{fig:optcond}
\end{figure}

The spectral function of pairs of distinguishable dopants in the $t-J$ model can be decomoposed into the sum of the spectra of symmetric (bosonic) and antisymmetric (fermionic) under exchange pairs of holes in the $t-J$ model,
\begin{equation}
A_{m_4}^{(\text{dist}-t-J)}(k,\omega) = A_{m_4}^{(f-t-J)}(k,\omega) + A_{m_4}^{(b-t-J)}(k,\omega).
\label{eq:A}
\end{equation}
We can therefore relate the different features in the excitonic spectral function to an emergent exchange symmetry of two dopants that comprise an exciton. This leads to two decoupled sectors corresponding to bosonic \cite{Homeier2024} and fermionic holes in the effective $t-J$ model. Using this understanding, we find for example that the excitonic peak observed in the optical conductivity, Fig.~\ref{fig:optcond}, is purely bosonic in nature (symmetric under exchange), since pairs of fermionic holes with $p$-wave symmetry do not have any spectral weight at momentum $\mathbf{k} = (0,0)$. Overall we find that the combination of bosonic and fermionic exchange symmetry sectors captures the excitonic spectra in the Fermi-Hubbard model at strong coupling $U\gg t$ remarkably well. 

Our results demonstrate that measurements of excitonic properties enable direct insights into the existence as well as properties of pairs of charge carriers and their excitations. Next we discuss signatures and measurements of excitons.


\section{Survey of solid state experiments}
In solid state experiments, excitons can be probed through different approaches: (i) optical measurements essentially access the current response to an electrical field, and thus probe current-current correlations; (ii) through Coulomb interaction processes, electrons can be excited from the lower to the upper Hubbard band in experiments such as angle-resolved electron energy loss spectroscopy (EELS) \cite{Schnatterly1979,Egerton2009}; (iii) in resonant inelastic x-ray spectroscopy (RIXS), Hubbard excitons can be created through interaction processes induced by a core hole \cite{Ament2011}.

Probes like the optical conductivity access excitons, as the application of the current operator along $\mu=x,y$,
\begin{equation}
\hat{j}_{\vec{l}}^{\mu} = i \sum_\sigma \left( \hat{c}_{\vec{l},\sigma}^\dagger \hat{c}_{\vec{l}+\vec{e_\mu},\sigma} -  \hat{c}_{\vec{l}+\vec{e_\mu},\sigma}^\dagger \hat{c}_{\vec{l},\sigma} \right),
\label{eqDefCurrent}
\end{equation}
can directly create an exciton with $p$-wave ($m_4=1$) symmetry. 
The existence of the bound state of doublon and hole manifests itself through a well-defined (excitonic) peak below a continuum of particle-hole excitations, i.e. below the upper Hubbard band. 
Remnants of an excitonic peak have for example been used to determine the nearest neighbor interaction strength $V$ in one-dimensional Mott insulators \cite{Jeckelmann2003,Wall2011,Mitrano2014}. 
The doublon-hole binding energy has been indirectly accessed in optical spectroscopy on three different undoped cuprate compounds, where potential evidence for an increase of the binding energy with $J$ has been found \cite{Terashige2019}.
The interplay between long-range as well as local antiferromagnetic spin correlations and exciton properties can be studied, for example, through pump-probe or pump-push-probe experiments with different delay times, thus heating up the spin background to different degrees \cite{Eckstein2016,Alpichshev2017}. Numerically, the effect of the spin background on excitonic properties has been studied by tuning the on-site Coulomb interaction $U/t$ and by the introduction of -- potentially frustrating -- next nearest neighbor hoppings $t'$ in simulations of the optical conductivity in the Fermi-Hubbard model using matrix product states \cite{Shinjo2021}.

Apart from optical probes, excitonic properties can also be accessed through EELS \cite{Wang1996,Zhang1998,Neudert1997,Neudert1998,Kuzian2003}, where the dynamic charge susceptibility is measured \cite{Abbamonte2024}. Nowadays, high momentum and frequency resolution are available in EELS experiments \cite{Vig2017,Abbamonte2024}. The dispersion of excitons can thus be accessed by tracking features close in energy to the Mott gap \cite{Egerton2009}. The coupling of EELS is not restricted to optically allowed transitions, potentially allowing to probe also excitons with $s$- or $d$-wave symmetry. 

RIXS provides another route to obtaining momentum- and frequency resolved spectra. 
In Cu K-edge RIXS, an electron is excited from the 1s to the 4p orbital. In the intermediate states, the interaction with the 1s-core hole leads to excitations in the 3d electron system across the Mott gap of the effective Hubbard model, thus creating a Hubbard exciton  \cite{Ament2011}. Finally, the 4p electron goes back to the 1s orbital under emission of a photon \cite{Tsutsui1999,Hasan2000,Wrobel2002,Tsutsui2003}. 
The created hole forms a Zhang-Rice singlet \cite{Zhang1988}, which carries angular momentum $m_4=2$, and an electron is excited onto the $d_{x^2-y^2}$ orbital of the neighboring Cu site, occupying the upper Hubbard band, and carrying $m_4=2$ \cite{Zhang1998,Hasan2000}. 
As the RIXS process involves two optical transitions, creation and recombination of the core hole, the total selection rule for the angular momentum is $\Delta l = -2,0,+2$. 
For an overall $\Delta l =2$ process, the resulting exciton has a $d$-wave structure. For $\Delta l =0$, an $s$-wave exciton is created. 
In the latter case, the doublon and the hole are spatially closer to each other. As a consequence in cuprates, we expect the $m_4=0$ exciton to be shifted to lower energies by the non-local Coulomb interaction and strongly broadened by the radiative decay, both of which is not taken into account in our numerical simulations. The signatures of the $m_4=0$ exciton in RIXS experiments on cuprates are hence expected to be less salient and at lower energy.
For a $d$-wave exciton, the amplitude to find the doublon and the hole localized around the same Cu site must be zero. The constitutents are thus more spread out and the Coulomb attraction between the doublon and the hole is smaller and radiative decay is potentially suppressed. Spectral features of $d$-wave excitons are thus expected to remain sharper, and a quantitative comparison of experimental data to $d$-wave excitons in the Hubbard model should work better than for $s$-wave excitons.

See Appendix~\ref{sec:solid} for further discussion of Hubbard Mott exciton probes in solid state experiments.


\section{Excitons in quantum simulators}
Properties of the clean Fermi-Hubbard model can be probed using quantum simulation experiments. These systems have the advantage of directly realizing the desired model without additional degrees of freedom such as phonons, which potentially suppress excitonic peaks in experiments on real materials \cite{Shinjo2021}. 
Similar to the relaxation process in ultrafast spectroscopy experiments on solids, the decay rates of doublons created through lattice modulation have been studied in the three-dimensional Hubbard model using cold atoms in optical lattices \cite{Strohmaier2010,Sensarma2010}. 
Recent advances in cold atom experiments enable the clean, well controlled realization of the two-dimensional Hubbard model at temperatures down to $T/J \approx 0.6$ with variable doping \cite{Mazurenko2017,Koepsell2021,Hartke2023}; see e.g.~\cite{Bohrdt2021_review} for a recent review of this field. 

As discussed in earlier work \cite{Joerdens2008,Sensarma2009,Strohmaier2010,Sensarma2010}, excitons can be created through global lattice modulation, where the depth of the optical lattice, and thus effectively the ratio of tunneling to interaction strength, is changed periodically in time. In the linear response regime, the number of doublons and holes can simply be counted in the routinely performed quantum projective measurements in order to obtain the zero momentum $s$-wave exciton spectrum, as shown in Fig.~\ref{fig:spectrum_k00}. This is analogous to the optical conductivity, although the latter involves couplings to the current operator and probes zero momentum $p$-wave excitons. The optical conductivity can be measured in ultracold atom experiments by lattice shaking~\cite{Tokuno2011,Wu2015,Anderson2019}. 

To measure zero momentum $d$-wave exciton spectra, we propose the use of independent intensity modulations of the optical lattices along $x$- and $y$-directions which are $\pi$ out-of-phase. This allows to transfer angular momentum to the system, as required for creating $d$-wave excitons. As shown in Fig.~\ref{fig:spectrum_k00}, the $d$-wave spectrum differs significantly from the $s$-wave exciton spectrum at $\mathbf{k}=(0,0)$. In particular, the shape and energies associated with the onset of spectral weight around $\omega - U = 0$ differ strongly.

Furthermore we propose to use Raman spectroscopy to measure momentum resolved exciton spectra. The momentum difference $\mathbf{k}$ and detuning $\omega$ of the Raman lasers directly allows to control frequency and momentum of the excited excitons. Since the Raman modulation affects the on-site potentials of the optical lattice, it couples directly to the current operator Eq.~\eqref{eqDefCurrent} and thus probes $p$-wave excitons only. In order to extract momentum resolved $s$- and $d$-wave spectra, a Raman setup producing a beat-note modulating the bonds of the optical lattice can be implemented. The latter couples directly to the kinetic energy, as in the case of global intensity modulation~\cite{Sensarma2009}; by controlling the relative phase of modulations along $x$- and $y$-directions, $s$- and $d$-wave spectra can be addressed individually. 

See Appendix~\ref{sec:coldAtoms} for a more detailed discussion of the exciton probes in quantum simulators summarized here.

\section{Model}
In this work, we study excitonic spectra in the Fermi-Hubbard model, 
\begin{equation}
\H_{FH} = - t \sum_{\ij} \sum_\sigma \l  \cd_{\vec{i},\sigma} \c_{\vec{j},\sigma} + \hc \r 
+  U \sum_{i} \hat{n}_{\vec{i},\uparrow} \hat{n}_{\vec{i},\downarrow},
\label{eq:FHHamiltonian}
\end{equation}
where $\hat{c}_{\vec{j},\sigma}^{(\dagger)}$ and $\hat{n}_{\vec{j},\sigma}$ denote fermionic annihilation (creation) and density operators, respectively. In our numerical matrix product state simulations, we consider a 40 site long, four-legged cylinder.

We define an operator $\hat{T}_{\pm \mu}(\vec{i},\sigma)$ that moves an electron with spin $\sigma$ from site $\vec{j} = \vec{i}\pm \vec{e}_\mu$, $\mu=x,y$ to site $\vec{i}$:
\begin{equation}
    \hat{T}_{\pm \mu}(\vec{i},\sigma) =     \cd_{\vec{i},\sigma} \c_{\vec{i }\pm \vec{e}_\mu,\sigma}.
    \label{eq:DefT}
\end{equation}
Applying this operator to a Mott insulator yields a doublon hole pair. We can now create an exciton on the bonds adjacent to site $\vec{i}$ with discrete $C_4$ angular momentum $m_4=0,1,2,3$ as $\hat{R}_{m_4}^x + \hat{R}_{m_4}^y$ with
\begin{equation}
    \hat{R}_{m_4}^\mu(\vec{i}) =\sum_\sigma \sum_{\tau=\pm} e^{i  m_4 \varphi_{-\tau \vec{e}_\mu}}   \hat{T}_{\tau \mu}(\vec{i},\sigma),
    \label{eq:DefR}
\end{equation}
with $\varphi_{\vec{r}} = {\rm arg}(\vec{r})$ the polar angle of $\vec{r}$.

Based on this operator, we now consider the excitonic Green's function
\begin{equation}
\mathcal{G}_{\rm exc}^{(m_4)}(\vec{k},t) = \theta(t)  \bra{\Psi_0} \sum_\mu \hat{R}_{m_4}^{ \mu,\dagger}(\vec{k},t) \sum_{\mu'}\hat{R}_{m_4}^{\mu'}(\vec{k},0) \ket{\Psi_0},
\label{eq:rotationalGF}
\end{equation}
which we calculate using time-dependent matrix product states \cite{Kjall2013,Zaletel2015,Paeckel2019} on top of the ground state of the half-filled Hubbard model $\ket{\Psi_0}$; see Appendix~\ref{sec:MPSconvergence} for details. The corresponding excitonic spectrum, $- \pi^{-1} {\rm Im} \mathcal{G}_{\rm exc}^{(m_4)}(\vec{k},\omega)$, in Lehmann representation is
\begin{equation}
 A_{\rm exc}^{(m_4)}(\vec{k},\omega) =  \sum_{n} \delta \l \omega-E_n+E_0 \r  | \bra{\Psi_n}\sum_\mu \hat{R}_{m_4}^\mu(\vec{k})  \ket{\Psi_0} |^2,
 \label{eq:DefArot}
\end{equation}
where $\ket{\Psi_n}$ ($E_n$)  are the eigenstates (eigenenergies) of the Fermi-Hubbard model at half-filling.   

\section{Optical conductivity}\label{sec:OptCond}
The excitonic spectrum is closely related to the optical conductivity, which is defined as
\begin{equation}
\sigma^{\mu \mu} (\omega) = \int_0^\infty dt~e^{i\omega t} C^{\mu\mu}(t)
\label{eq:Defsigma}
\end{equation}
with
\begin{multline}
C^{\mu \mu} (t) =- \sum_{\vec{i}} \bra{\Psi_0} \hat{R}_{m_4=1}^{\mu,\dagger}(\vec{i},t) 
 \hat{R}_{m_4=1}^\mu(\vec{0},0)  \ket{\Psi_0} e^{iE_0 t},
 \label{eq:Defsigmat}
\end{multline}
and corresponds to the current response function, hence $m_4=1$. 

The optical conductivity, $\sigma^{\mu \mu}(\omega)$ is routinely measured in materials such as the cuprates \cite{Uchida1991}. 
Fig.~\ref{fig:optcond} shows the optical conductivity, which exhibits two main features: a low energy peak at energies $\approx J$, corresponding to spin-wave excitations; and a second peak at energies just below the broad continuum constituting the upper Hubbard band. This latter, high-energy feature in $\text{Re}\sigma(\omega)$ at the Mott gap edge is the lowest-energy exciton peak, which is well resolved in our numerics, Fig.~\ref{fig:optcond}, consistent with earlier numerical results using exact diagonalization \cite{Tohyama2005} and DMFT \cite{Taranto2012}. 

Next we compare the optical conductivity of the Hubbard model to predictions by a $t-J$ model of interacting doublons and holes that will be further specified below. This model can capture the physics associated with the upper Hubbard band. 
Intriguingly, the qualitative structure of the optical conductivity in the Hubbard model at energies around $U$ is well reproduced by the corresponding probe in the $t-J$ model, where the optical conductivity measures the linear response to the creation of a pair of distinguishable dopants, see Sec. \ref{sec:tJ}. 
Due to symmetry considerations, the corresponding response function for indistinguishable holes which are anti-symmetric under exchange (fermionic statistics) is strictly zero, as this excitation has $p$-wave character and is measured at momentum $\mathbf{k}=(0,0)$. Thus,  we conclude that the excitonic peak in the optical conductivity, which has been observed numerically in the Hubbard model~\cite{Dagotto1992,Onodera2004,Tohyama2005,Taranto2012,Shinjo2021} and in experiments on cuprate materials \cite{Uchida1991,Terashige2019}, corresponds to indistinguishable dopants which are symmetric under exchange (bosonic statistics).  

\section{$t-J$ model of excitons}\label{sec:tJ}
In the limit of large $U/t$, the Fermi-Hubbard model can be approximated by the $t-J$ model, 
\begin{multline}
\H_{t-J} = - t ~ \hat{\mathcal{P}} \sum_{\ij} \sum_\sigma \l  \cd_{\vec{i},\sigma} \c_{\vec{j},\sigma} + \hc \r \hat{\mathcal{P}} + \\
+ J \sum_{\ij} \hat{\vec{S}}_{\vec{i}} \cdot \hat{\vec{S}}_{\vec{j}} - \frac{J}{4} \sum_{\ij} \hat{n}_{\vec{i}} \hat{n}_{\vec{j}},
\end{multline}
where $\hat{\mathcal{P}}$ projects to the subspace with maximum single occupancy per site; $\hat{\vec{S}}_{\vec{j}}$ and $\hat{n}_{\vec{j}}$ denote the on-site spin and density operators, respectively. 
We consider three different versions of the $t-J$ model, each with exactly two dopants of different character:
\begin{itemize}
\item[(i)] dopants which are symmetric under exchange 
\item[(ii)] dopants which are anti-symmetric under exchange 
\item[(iii)] equal but distinguishable dopants.
\end{itemize}
The latter case is the combination of cases (i) and (ii). The eigenstates of the Hamiltonian in case (iii) have a well-defined parity under particle exchange, namely $+1$ (bosonic statistics, case (i)) and $-1$ (fermionic statistics, case (ii)).
Properties of two distinguishable dopants, such as peaks in the spectral function, can thus be decomposed into a sum of a symmetric and anti-symmetric contribution, see Eq.~\eqref{eq:A}. 

In all cases, we consider two dopants, which are governed by the same terms in the Hamiltonian, but in case (iii) can be distinguished e.g. through an explicit label, as realized if a doublon and a hole excitation are created in the Fermi-Hubbard model. We can make this more explicit by introducing spinless fermionic chargon operators $\hat{h}_j$, describing holons, and $\hat{d}_j$, describing doublons, and Schwinger bosons $\hat{b}_{j\sigma}$, such that we can write the electron operator as 
\begin{equation}
\hat{c}_{j\sigma}^\dagger = (-1)^{j_x+j_y}\hat{h}_j \hat{b}_{j\sigma}^\dagger +  \hat{d}_j^\dagger \hat{b}_{j\bar{\sigma}},
\end{equation}
see also \cite{Huang2023}. 
In the first term, we have applied a gauge transformation to obtain hopping terms with the same sign for $\hat{h}$ and $\hat{d}$ operators in the Hamiltonian below. 
The single occupancy constraint due to the projection operator $\hat{\mathcal{P}}$ can then be written as $\sum_\sigma \hat{b}_{j\sigma}^\dagger \hat{b}_{j\sigma} + \hat{h}_j^\dagger\hat{h}_j + \hat{d}_j^\dagger \hat{d}_j = 1$ and the Hamiltonian for the doublon and the hole becomes
\begin{align}
\begin{split}
\H_{t-J} = &- t ~  \sum_{\ij} \sum_\sigma \l  \hat{h}_{\vec{j}}^\dagger \hat{b}_{\vec{i}\sigma}^\dagger \hat{b}_{\vec{j}\sigma} \hat{h}_{\vec{i}} + \hc  \r  + \\
&- t ~  \sum_{\ij} \sum_\sigma \l  \hat{d}_{\vec{i}}^\dagger \hat{b}_{\vec{i}\sigma}^\dagger \hat{b}_{\vec{j}\sigma} \hat{d}_{\vec{j}} + \hc  \r  + \\
&+ J \sum_{\ij} \hat{\vec{S}}_{\vec{i}} \cdot \hat{\vec{S}}_{\vec{j}} - \frac{J}{4} \sum_{\ij} \hat{n}_{\vec{i}} \hat{n}_{\vec{j}} + U\sum_{\vec{i}} \hat{d}_{\vec{i}}^\dagger  \hat{d}_{\vec{i}}.
\end{split} 
\end{align}

In the following, we compare the excitonic spectra evaluated in the Fermi-Hubbard model to \emph{pair} spectra in the $t-J$ model. Excitonic spectra correspond to the response function with an excitation where a fermion hops from one site to its neighbor as described by the operator $\hat{T}_{\pm \mu}(\vec{i},\sigma)$. 

For two distinguishable dopants in the $t-J$ model, corresponding to doublon and hole, we consider the same excitation, which we can now express by
\begin{equation}
    \hat{T}_{\tau \mu}(\vec{i},\sigma) =       \hat{\Delta}_{\tau \mu}(\vec{i}) =   \hat{d}_{\vec{i}}^\dagger \hat{h}_{\vec{i} + \tau \vec{e}_\mu}^\dagger \nicefrac{1}{\sqrt{2}} \sum_\sigma (-1)^\sigma \hat{b}_{\vec{i}\sigma} \hat{b}_{\vec{i} + \tau \vec{e}_\mu,\sigma}.
\end{equation}
The excitation operator, corresponding to Eq.~\eqref{eq:DefR} in the Hubbard model, then becomes
\begin{equation}
\hat{R}_{m_4}^\mu(\vec{i}) = \sum_{\tau=\pm} e^{im_4\phi_{\tau\vec{e}_\mu}} \hat{T}_{\tau \mu}(\vec{i},\sigma).
\end{equation}

In the case of the $t-J$ models with (anti-)symmetric statistics, corresponding to cases (i), (ii) above, we replace the excitation $\hat{T}_{\pm \mu}(\vec{i},\sigma)$ in equations \eqref{eq:DefR} to \eqref{eq:Defsigmat}  by 
\begin{equation}
    \hat{\Delta}_{\pm \mu}(\vec{i},\sigma) =     \c_{\vec{i},\sigma} \c_{\vec{i }\pm \vec{e}_\mu,\bar{\sigma}}
    \label{eq:DefDelta}
\end{equation}
with fermionic (bosonic) $\hat{c}_{\vec{i},\sigma}$, see also \cite{Grusdt2022,Bohrdt2023} for a detailed analysis of pair spectra in the $t-J$ model. 

As can be seen in Fig.~\ref{fig:optcond}, the optical conductivity calculated for the Fermi-Hubbard model agrees very well with the corresponding response function in the $t-J$ model with distinguishable dopants. In particular, the $t-J$ model also features a well-defined excitonic peak at energies below $U$.  The low-energy peak, corresponding to spin excitations in the Hubbard model, does not appear in the $t-J$ model, as the perturbation in the latter case does not include the corresponding processes. 



\begin{figure}
\centering
\includegraphics[width=0.99\linewidth]{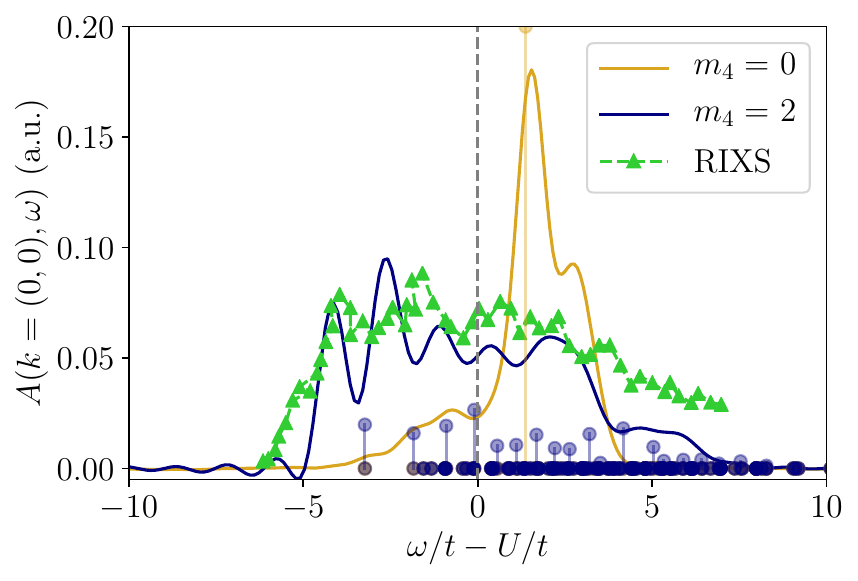} 
\caption{
\textbf{Exciton spectra} at momentum $\mathbf{k}=(0,0)$ in the Fermi-Hubbard model for $U/t=12$ for an $s$-wave ($m_4=0$, orange) and a $d$-wave ($m_4=2$, blue) excitation. Green data points denote the Cu K-edge RIXS measurements extracted from Ref.~\cite{Ellis2008}. Opaque markers denote peak positions and spectral weight (indicated by height) predicted by the geometric string theory for two fermionic dopants.
}
\label{fig:spectrum_k00}
\end{figure}

\section{Ro-vibrational excitations}\label{sec:Rovibrations}
A natural extension of the optical conductivity to other angular momenta is the excitonic spectrum, as defined in \eqref{eq:DefArot}, at momentum $\mathbf{k}=(0,0)$, which can be probed in quantum simulation experiments for clean model systems.  Note that in the excitonic spectrum, we sum over both directions $\mu=x,y$, whereas in the optical conductivity, only one direction is considered \cite{Tohyama2006}. 

In Fig.~\ref{fig:spectrum_k00}, we show the excitonic spectrum at $\mathbf{k}=(0,0)$ for an $s$-wave as well as a $d$-wave excitation, revealing a striking difference between the two angular momenta: in the former case, there is almost no spectral weight below an energy of $\omega \approx U$, whereas in the latter case, well-defined excitations at energies $\Delta \omega \approx 4 t$ below $U$ are visible. 
In the $d$-wave spectrum, we observe several peaks across an energy window of approximately $8t$, which we interpret as excitations of the string binding the doublon and hole together. 

The (geometric) string theory comprises a theoretical description of a single dopant, where the interplay of antiferromagnetic spin exchange and kinetic energy of the dopant leads to the formation of a string of displaced spins which binds the chargon (charge $1$, spin $0$ excitation) to the spinon (charge $0$, spin $1/2$ excitation) \cite{Bulaevski1968,Grusdt2018tJz,Grusdt2019}.
The same theoretical framework has also been applied to the analysis of two dopants, which can be similarly bound to each other through a string of displaced spins \cite{Shraiman1988,Grusdt2022,Bohrdt2022_bilayer}

In the single particle spectral function obtained in linear spin wave theory \cite{Martinez1991,Liu1992}, a multi-peak structure has been predicted and interpreted as string excitations \cite{Bulaevski1968,Liu1992,Beran1996,Manousakis2007}. In large scale numerical calculations of the $t-J$ model \cite{Brunner2000,Mishchenko2001,Bohrdt2020_ARPES}, however, only the first vibrational peak has been observed. In contrast to the single particle spectral function, which probes an individual magnetic polaron, we are here considering two dopants (doublon and hole). In the geometric string theory, the effective hopping relative to the spin-exchange is thus increased \cite{Grusdt2022}, yielding more stable string excitations. 
These string excitations can therefore be observed numerically, as shown in Fig.~\ref{fig:spectrum_k00}, consistent with string theory predictions, see Appendix~\ref{sec:strings}. 

Intriguingly, Cu K-edge RIXS measurements on $\text{La}_2\text{CuO}_4$ \cite{Ellis2008}, green triangles in Fig.~\ref{fig:spectrum_k00}, exhibit a very similar multi-peak structure. In comparing the RIXS data to our numerical results, we have assumed $t=350$meV as determined from ARPES measurements on $\text{La}_2\text{CuO}_4$. We have used one global fitting parameter in all comparisons of RIXS data from \cite{Ellis2008}, consisting in an overall energy shift of $\Delta E/t = 10.2$, corresponding to the expected scale of the Hubbard interaction $U/t$. Remarkably, the frequency dependence at all available momenta agrees well with our numerical results for the $d$-wave exciton spectra, see Fig.~\ref{fig:RIXS_allk}.

A similar multi-peak structure has been observed in the optical conductivity obtained with DMFT \cite{Taranto2012}, and was interpreted as a signature of magnetic polaron formation. As shown in Sec.~\ref{sec:OptCond}, the optical conductivity serves as a probe of Hubbard excitons, which can be related to pairs of dopants. We thus argue that the higher excited peaks observed in \cite{Taranto2012} correspond to excitations of the string binding the two dopants (doublon and hole) together instead of being related to individual, unbound dopants. Intriguingly, broad peaks which can be interpreted as signatures of such higher string excitations have also been observed in optical conductivity measurements on $\text{La}_2\text{CuO}_4$ and $\text{Sr}_2\text{CuO}_2\text{Cl}_2$ \cite{Terashige2019}.

\section{Exciton dispersion}
\begin{figure}
\centering
\includegraphics[width=0.95\linewidth]{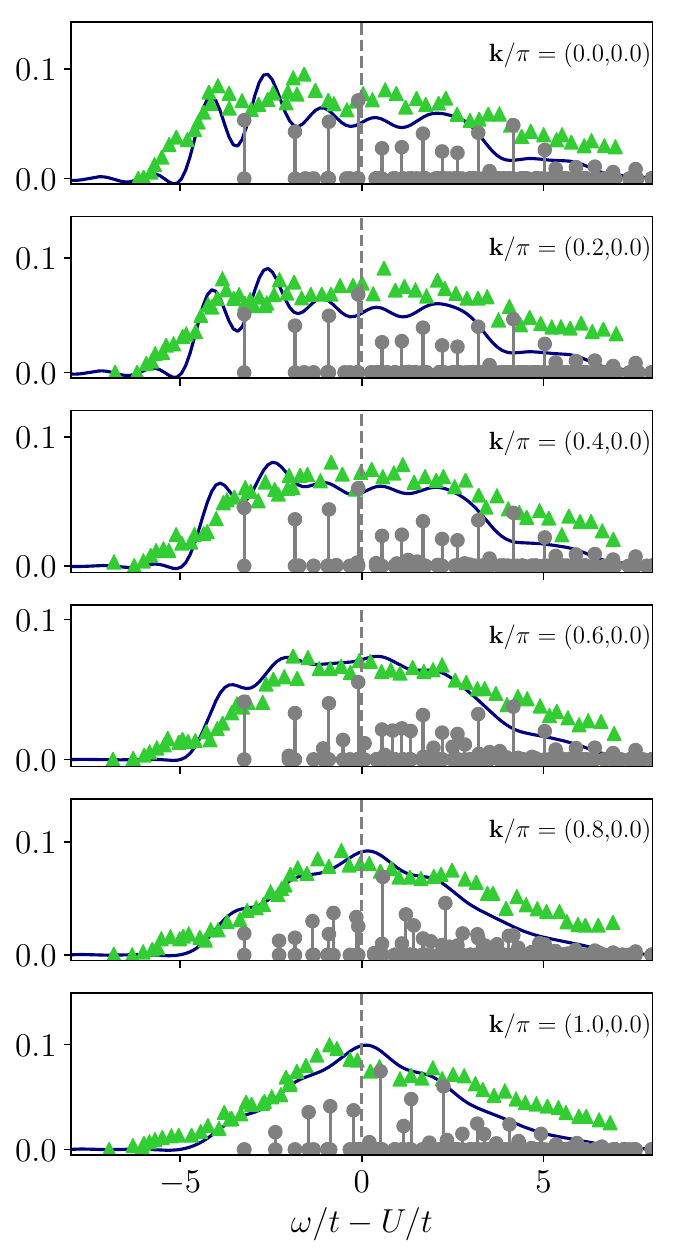} 
\caption{
\textbf{Exciton spectra and RIXS data} at different momenta $\vec{k}$ as indicated. Numerical simulation of exciton spectra in the Hubbard model for $U/t=12$ and $d$-wave ($m_4=2$) excitation. RIXS data extracted from \cite{Ellis2008}, assuming $t=350$meV. An overall energy shift of $\Delta \omega = 10.2$ was applied to the RIXS data. The weights have been rescaled to yield the same maximum weight between RIXS and numerical results for each momentum cut. Gray markers denote peak positions and spectral weight predicted by the geometric string theory for two fermionic dopants for $m_4=2$ and $t/J=3$.
}
\label{fig:RIXS_allk}
\end{figure}

Beyond the zero momentum probes discussed so far, the exciton spectrum as defined in Eq.~\eqref{eq:DefArot} allows to additionally investigate their momentum dependence, see e.g. Fig.~\ref{fig:RIXS_allk}. 
In particular, by considering the momentum resolved exciton spectrum from our MPS simulations, we can directly gain insights into the dispersion relation of Hubbard excitons.  
Similarly, the momentum resolution can also be probed in quantum gas microscopy experiments directly realizing the Fermi-Hubbard model, as well as in experiments on quantum materials, using e.g. EELS or RIXS. As shown in Ref.~\cite{Kim2009}, the results from EELS and RIXS on undoped cuprate compounds exhibit dispersive peaks and agree qualitatively well.

In Fig.~\ref{fig:spectrum_numericsRIXS}(a), we show the momentum $k_x$ and energy resolved exciton spectra for an angular momentum $m_4=2$, i.e. a $d$-wave excitation. We consider energies of a few hoppings $t$ below the Hubbard interaction energy $U$ and find a well defined peak in the spectrum for all momenta $k_x$. This low energy peak exhibits a comparably weak, but clearly visible dispersion on the order of one hopping $t$. A second peak with the same dispersion is visible at slightly higher energies, which can be interpreted as an excited internal (string) state of the exciton.  

We compare the numerical results to RIXS measurements on two different undoped cuprate materials, in particular $\text{Ca}_2\text{CuO}_2\text{Cl}_2$ \cite{Hasan2000} (squares) and $\text{La}_2\text{CuO}_4$ (up triangles from \cite{Ellis2008}, stars from \cite{Collart2006}, down triangles from \cite{Kim2002}). 
We assume a hopping strength of $t=350\text{meV}$, as determined e.g. by independent ARPES experiments \cite{Wells1995},  and allow for a global frequency shift. Notably, the dispersion of the lowest branch in our numerical simulations agrees well with almost all measurements for $0\leq k_x\lesssim \pi/2$. For larger momenta, the spectral weight at low energies decreases, and the experimental results show a larger spread. The resolution in Ref.~\cite{Kim2002} is significantly lower compared to the other experimental data, which potentially explains the discrepancy. According to our numerical analysis, \cite{Kim2002} mainly captures the first excited branch, which has higher spectral weight compared to the excitonic ground state. 

We numerically find the ro-vibrational string excitations discussed in Sec.~\ref{sec:Rovibrations} also at momenta away from $\vec{k}=(0,0)$, with an overall momentum and frequency dependence that agrees remarkably well with RIXS experiments on $\text{La}_2\text{CuO}_4$ \cite{Ellis2008}, as shown in Fig.~\ref{fig:RIXS_allk}.


 \begin{figure}
\centering
\includegraphics[width=0.95\linewidth]{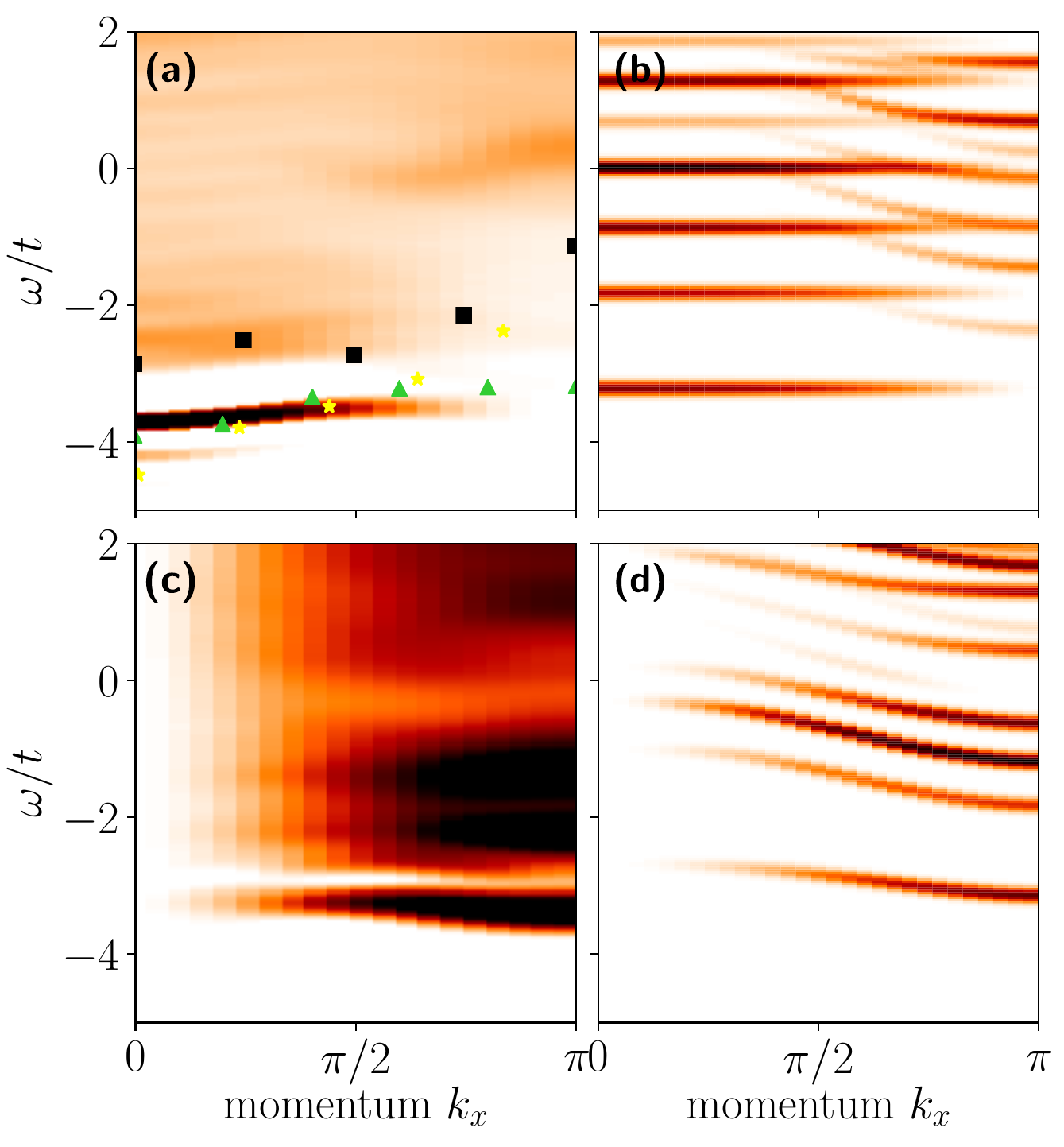} 
\caption{
\textbf{Pair spectra} with momentum resolution in $x$-direction at $k_y=0$ for two dopants with (a), (b) anti-symmetric and (c), (d) symmetric exchange statistics in the $t-J$ model for $t/J=3$ for a $d$-wave ($m_4=2$) excitation.  For the anti-symmetric dopants, we show the peak positions as extracted from RIXS measurements on $\text{Ca}_2\text{CuO}_2\text{Cl}_2$ \cite{Hasan2000} (squares), $\text{La}_2\text{CuO}_4$ (up triangles from \cite{Ellis2008}, stars from \cite{Collart2006}, down triangles from \cite{Kim2002}) assuming $t=350\text{meV}$ and using an energy offset of $\Delta E=10.2 t$, consistent with ARPES results \cite{Wells1995}. (a), (c) are time-dependent matrix product state simulations of the (anti-)symmetric $t-J$ model, whereas (b), (d) are the spectral functions obtained from geometric string theory calculations \cite{Grusdt2022} (lines broadened for illustration).
}
\label{fig:bosons_fermions}
\end{figure}

In Fig. \ref{fig:spectrum_numericsRIXS}(b), we also compare the exciton spectrum to the corresponding pair spectral function in the $t-J$ model with distinguishable dopants. Qualitatively, similar features can be found in both cases, in particular a pronounced peak at low energies around $\mathbf{k}=(0,0)$, with a dispersion of approximately one hopping $t$ and vanishing spectral weight around $\mathbf{k}=(\pi,0)$. 
Similarly, for an $s$-wave excitation, a strongly dispersive high-energy feature, starting at frequencies $\omega/t \approx 2$ above $U$ at $\mathbf{k}=(0,0)$, is visible for the Hubbard exciton as well as for two distinguishable dopants in the $t-J$ model, see Appendix~\ref{sec:MoreSpectra}. 

Since the spectral function for two distinguishable dopants is composed of the spectra for two indistinguishable dopants (i) symmetric and (ii) anti-symmetric under exchange, see Eq.~\ref{eq:A}, we can associate various features in the Fermi-Hubbard exciton spectrum with either symmetric or anti-symmetric pairs. 
To this end, in Fig.~\ref{fig:bosons_fermions}, we separately consider the pair spectral function for the $t-J$ model with (anti-)symmetric dopants, as evaluated from MPS simulations for the microscopic model (left column), as well as for the geometric string theory (right column) \cite{Grusdt2022}. 
Qualitatively, we find good agreement between the numerical simulation of the microscopic model, and the semi-analytical geometric string theory. In particular, the low energy peaks found in the numerically calculated spectra can be identified with bands found in the geometric string theory, enabling a physical understanding of the different observed features. Moreover, the comparison to the Hubbard exciton spectra as well as the RIXS data shows that the observed lowest peak with dispersion $\approx t$ for $0\leq k_x \leq \pi/2$ is due to the anti-symmetric (fermionic) component. 

The similarities between the excitonic spectra and the two dopant spectra in the $t-J$ model show that experimental observations of excitonic properties enable insights into the properties of pairs of charge carriers, which are otherwise challenging to access. 
Note that the binding energy of the exciton, and equivalently the pair, is not completely determined by the position of the peak in the excitonic or two-dopant spectrum: in order to obtain the binding energy, the frequency of the two-dopant peak has to be compared to twice the energy of the peak position in the single dopant spectrum. 

It is instructive to compare exciton dispersions to single particle dispersions accessible through angle-resolved photoemission spectroscopy (ARPES). From a theoretical perspective, the propagation of a single dopant is due to spin exchange processes, yielding a dispersion on the order of the superexchange $J$ \cite{Kane1989,Sachdev1989,Beran1996,Grusdt2018tJz,Grusdt2019}, in agreement with ARPES measurements on cuprates \cite{Wells1995}. 
The excitonic branches observed in EELS and RIXS experiments exhibit a substantially larger dispersion than the one-particle excitations \cite{Wang1996,Neudert1997,Fink2001,Moskvin2002}, in particular along the diagonal in momentum space from $\mathbf{k}=(0,0)$ to $(\pi,\pi)$, with a bandwidth that scales with the hopping amplitude $t$. Experimental evidence thus indicates a significantly lower effective mass of the exciton compared to a single dopant, implying strong correlations between the doublon and the hole constituting the exciton. This dichotomy was previously predicted for pairs of two fermionic holes~\cite{Bohrdt2023}.

Our present analysis confirms these earlier predictions in the context of excitons. The dispersion extracted from RIXS data shown in Fig.~\ref{fig:spectrum_numericsRIXS} and Fig.~\ref{fig:bosons_fermions} along the $\mathbf{k}=(0,0)$ to $(\pi,0)$ direction agrees well with numerically calculated $d$-wave pair spectra, where the dispersion is largely determined by $J$ \cite{Bohrdt2023}. 
A strongly dispersive feature is numerically observed for $s$-wave pairs, which however is not clearly visible in the experimental RIXS data. 
In EELS experiments, a large-bandwidth feature is observed along the diagonal of the Brillouin zone, where finite system sizes inhibit a comparison to our numerical simulations. These experimental measurements are consistent with predictions by the geometric string theory for $p$-wave pairs however, see Appendix~\ref{sec:strings}.

\section{Summary and Outlook}
In this work we have calculated momentum and frequency resolved spectra of Hubbard excitons using time-dependent matrix product state simulations, yielding an unprecedented high resolution in momentum space. The features observed in the excitonic spectra can largely be identified with features found in spectra of two dopants in the $t-J$ model, thus enabling insights into the pairing mechanism in these strongly correlated models. The comparison with semi-analytical methods like the geometric string theory facilitates the interpretation of different peaks. 

We compare the excitonic spectra in the Hubbard model to spectra of two distinguishable dopants in the $t-J$ model. The latter can be exactly decomposed into a $t-J$ model with dopants (i) symmetric and (ii) anti-symmetric under exchange, corresponding to (i) bosonic and (ii) fermionic dopants. In the context of the doped Fermi-Hubbard model, the $t-J$ model with fermionic exchange statistics naturally arises and is typically studied in the literature. Using additional internal states in bosonic quantum simulators, e.g. Rydberg or molecule tweezer arrays, a bosonic version of the $t-J$ model with antiferromagnetic couplings can also be realized and probed in quantum simulation experiments \cite{Homeier2024}, enabling more detailed tests of our description of excitons in the future. More generally, such studies allow to explore the microscopic pairing mechanism of doped charges in quantum antiferromagnets, in which bound states involving strings of displaced spins -- as revealed here to underlie Hubbard Mott exciton formation -- may play a central role~\cite{homeier2023feshbach}.

The numerically obtained dispersion relation, as well as the entire frequency dependence for various momentum cuts, of $d$-wave excitons qualitatively agrees with Cu K-edge RIXS measurements on two different undoped cuprate compounds \cite{Hasan2000,Kim2002,Collart2006,Ellis2008}. 
Comparing the experimental measurements to our numerical results thus (i) enables an interpretation of the RIXS measurements in terms of the exciton spectra, on the level of a single-band Hubbard model, and (ii) strongly suggest that the excitons observed in these experiments have a $d$-wave symmetry, in agreement with theoretical expectations. 
Moreover, excitonic spectra can also be measured in quantum simulation experiments using cold atoms in optical lattices, for example through lattice modulation or Raman spectroscopy. The spectra of Hubbard excitons thus provide a valuable bridge between clean model systems as realized in quantum simulators and numerics, and real materials, while simultaneously allowing for insights into the properties of pairs of dopants in quantum antiferromagnets.  

Our numerical results in this work are limited to four-leg cylinders, and quantitative changes are expected when increasing the number of legs, as seen for the optical conductivity in Ref.~\cite{Shinjo2021}, where in the numerical results for six legs a more pronounced excitonic peak is observed compared to the four leg results. 
The limited system size in one direction inhibits the comparison of the numerically obtained exciton dispersion along the diagonal $(0,0)$ to $(\pi,\pi)$ cut, where a strongly dispersive feature has been observed both in EELS \cite{Wang1996,Moskvin2002} and RIXS \cite{Hasan2000,Ellis2008}, consistent with semi-analytical string theory results, see App.~\ref{sec:strings}. A comparison of the solid state measurements to results for the Fermi-Hubbard model in larger system sizes would thus be desirable.  

Intriguingly, EELS and RIXS measurements on doped cuprate compounds show a still well defined excitonic peak at 5 and 10 \% doping, which is however within the experimental resolution non-dispersive \cite{Schuster2009,Ellis2011}. The multi-peak structure observed in RIXS measurements on the Mott insulator $\text{La}_2\text{CuO}_4$ \cite{Collart2006,Ellis2008} as well as in our numerical results, see Fig.~\ref{fig:spectrum_k00} and \ref{fig:RIXS_allk}, is also observed at optimal doping in $\text{HgBa}_2\text{CuO}_{4+\delta}$ \cite{Lu2005}. 
These excited peaks at finite doping also become non-dispersive, which suggests a direct relation to the lowest energy excitonic peak, provided that disorder effects do not dominate the signal. Based on these measurements,  an interpretation in terms of the single-band Hubbard model up to optimal doping is suggestive.
These observations on cuprate materials in regimes which are numerically extremely challenging render the measurement of excitonic spectra in large, clean realizations of the Fermi-Hubbard model in quantum simulators highly desirable.  
Interesting directions for future research include the study of Hubbard excitons on other geometries, in particular non-bipartite lattices, such as the triangular lattice, and the square lattice with a nearest neighbor hopping $t'$, which has recently gained attention as it may help to stabilize a superconducting phase \cite{Xu2024}. 
\\


\textbf{Acknowledgements.--}
We thank Pit Bermes, Immanuel Bloch, Petar Bojovic, Antoine Georges, Isabella Gierz, Markus Greiner, Mohammad Hafezi, Timon Hilker, Lukas Homeier, Matteo Mitrano, and Ulrich Schollw\"ock for fruitful discussions. This research was funded by the Deutsche Forschungsgemeinschaft (DFG, German Research Foundation) under Germany's Excellence Strategy -- EXC-2111 -- 390814868, by the European Research Council (ERC) under the European Union’s Horizon 2020 research and innovation programm (Grant Agreement no 948141) — ERC Starting Grant SimUcQuam, by the ARO (grant number W911NF-20-1-0163), by the NSF through a grant for the Institute for Theoretical Atomic, Molecular, and Optical Physics at Harvard University and the Smithsonian Astrophysical Observatory, and by the SNSF project 200021\_212899,  support from the Swiss State Secretariat for Education, Research and Innovation (contract number UeM019-1), and the ARO grant number W911NF-20-1-0163.

%

\newpage
\appendix


\section{Solid state experiments}\label{sec:solid}

In general, the relaxation of a strongly correlated system after a perturbation is a complicated process. In quantum materials in particular, contributions from different degrees of freedom have to be disentangled.

\subsection{Optical probes}
Potential Hubbard exciton resonances in optical probes of different AFM Mott insulators are often reported as broad peaks overlapping with the continuum above the Mott gap \cite{Goessling2008,Gretarsson2013,Terashige2019,Shinjo2021}.

As the excitation in the case of the optical conductivity is given by the current operator, it only couples to excitons with a $p$-wave symmetry, and thus does not probe $s$- or $d$-wave pairing between electrons and holes. Early theoretical work pointed out that the excitonic bound state lowest in energy should be of $s$-wave symmetry and is thus optically forbidden \cite{Zhang1998,Tohyama2006}. 
Large-shift Raman scattering enables direct access to the symmetry of the photoexcited states, including for optically forbidden transitions \cite{Salamon1995,Tohyama2006}. 
Raman scattering is frequently used to probe magnetic properties of the system. However, if the energy of the incident photon is close to the Mott gap, it can also reveal excitonic properties \cite{Chubukov1995,Blumberg1995}.


Evidence for a low energy exciton bound state with $s$-wave symmetry is also seen for example in terahertz pump optical reflectivity probe spectroscopy on cuprate materials,  where the measured spectra are described by an effective three level model, consisting of the original undoped state without an exciton, as well as an additional $s$-wave or $p$-wave exciton \cite{Terashige2019}. Similarly, ultrafast terahertz conductivity measurements on the antiferromagnetic Mott insulator $\text{Sr}_2\text{IrO}_4$ are consistent with intra-excitonic transitions between states with $s$-, $d$-, and $p$-wave symmetry \cite{Mehio2023}.

\subsection{RIXS}\label{sec:RIXS}

While the details of the RIXS cross section and their relation to fundamental response functions are debated, there are strong arguments that the parallel-polarized RIXS signal can provide a probe for the charge structure factor, at least in terms of line shapes and the poles of the dynamical susceptibility, if not spectral weights \cite{Jia2016,Ament2011}.

\subsection{EELS: Comparison to string theory}\label{sec:strings}

\begin{figure}
\centering
\includegraphics[width=0.95\linewidth]{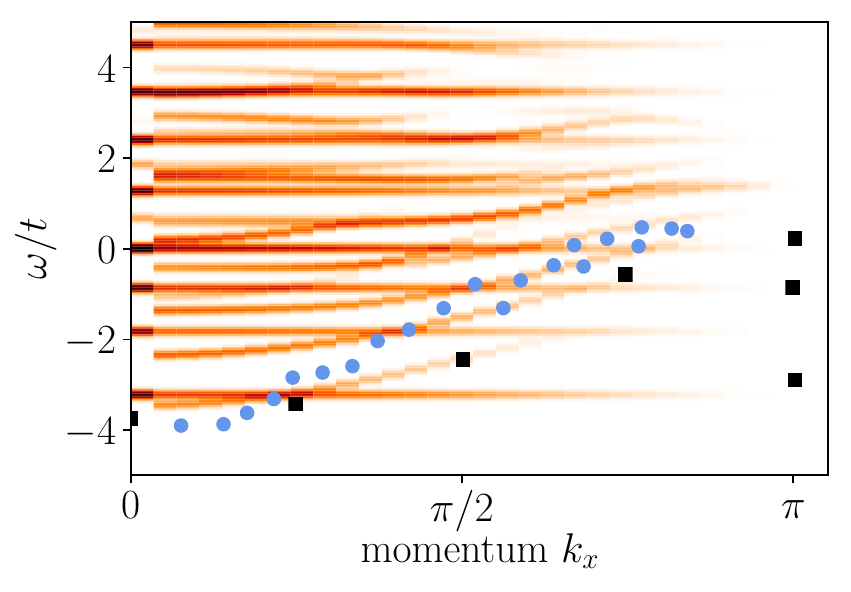} 
\caption{
\textbf{Geometric string theory results} for the diagonal cut $\vec{k}=(0,0)$ to $(\pi,\pi)$ for two distinguishable dopants in the $t-J$ model for $t/J=3$ for a $p$-wave ($m_4=1$) excitation, compared to EELS \cite{Wang1996} and RIXS \cite{Hasan2000} measurements.
}
\label{fig:k00_pipi_strings}
\end{figure}

As discussed in the main text, EELS can also be used to probe excitonic properties. Early EELS results on $\text{Sr}_2\text{CuO}_2\text{Cl}_2$ reported a highly anisotropic dispersion for the two cuts along $(0,0)$ to $(\pi,0)$ versus to $(\pi,\pi)$ through the Brillouin zone \cite{Wang1996}. In the main text, we compared the excitonic dispersion as obtained from our numerical simulations for $m_4=2$ to RIXS data for the first cut and obtained good agreement, qualitatively probing the comparably weakly dispersing pair observed in numerical and semi-analytical studies in \cite{Grusdt2022,Bohrdt2023}. While our numerical results are limited to four-legged cylinders and thus do not allow for a comparison along the diagonal cut through the Brillouin zone, the spectra of two distinguishable dopants can be calculated using the semi-analytical geometric string theory. In Fig.~\ref{fig:k00_pipi_strings}, we compare the resulting semi-analytical spectra at $m_4=1$ to the peak positions extracted from EELS measurements on $\text{Sr}_2\text{CuO}_2\text{Cl}_2$ \cite{Wang1996}. In both cases, a strongly dispersive feature, reminiscent of the 'light' pair in the $t-J$ model for an $s$-wave excitation ($m_4=0$) \cite{Bohrdt2023}, is observed.

\section{Probing exciton spectra with ultracold atoms}\label{sec:coldAtoms}
In this appendix we describe how exciton spectra can be conveniently measured in existing setups of ultracold fermions in optical lattices. The probes we describe neither require single-site nor single-atom resolution, and are suitable both for setups using quantum gas microscopes or bulk gases in two or three dimensions. For concreteness, we consider hypercupbic lattices only, although extensions to other lattices are straightforward.

In the following we consider different lattice modulations, which we assume to be sufficiently weak. On the one hand this allows to continue working in the lowest Bloch band of the underlying physics lattice and describe the effect of the lattice modulation on a tight-binding level. On the other hand, linear-response theory allows to relate a given lattice modulation with amplitude $\delta \mathcal{M}$ at frequency $\omega$ and momentum $\mathbf{k}$,
\begin{equation}
    \delta \H = \delta \mathcal{M} \sum_{\mathbf{r}} \cos(\mathbf{k} \cdot \mathbf{r} - \omega t) \hat{O}_{\mathbf{r}},
\end{equation}
to a spectral function in Lehmann representation,
\begin{equation}
    A_O(\omega,\mathbf{k}) = \sum_n | \bra{\psi_n} \hat{O}(\mathbf{k}) \ket{\psi_0}|^2 \delta(\omega-E_n+E_0);
\end{equation}
Namely, the rate $\Gamma(\mathbf{k},\omega)$ at which excitations of type $\hat{O}$ are created is proportional to
\begin{equation}
    \Gamma(\mathbf{k},\omega) \propto |\delta \mathcal{M}|^2 A_O(\omega,\mathbf{k}).
\end{equation}
Thereby counting of such excitations following a weak modulation provides a direct measurement of the spectral function $A_O$.

\subsection{Lattice intensity modulation}
The conceptually simplest modulation involves intensities of the primary laser beams defining the optical lattice. Assuming non-interfering beams along $x$- and $y$-directions~\cite{Bloch2008}, the modulated lattice potential reads
\begin{equation}
    V(\mathbf{r},t)= \sum_{\mu=x,y} \left( V_0 + \delta V \cos(\omega t - \varphi_\mu) \right) \cos^2 \left( \frac{\pi}{a_\mu} r_\mu \right),
\end{equation}
where the relative phase $\varphi_x-\varphi_y$ of modulations along $x$- and $y$-directions can be directly controlled; $a_\mu$ denotes the lattice constant.

On a tight-binding level, this generates a modulation of the kinetic energy, or tunneling terms, at zero momentum, $\mathbf{k}=(0,0)$,
\begin{equation}
    \delta \H  = \delta t \sum_{\mu=x,y} \cos(\omega t - \varphi_\mu) \sum_{\mathbf{j}} \sum_\sigma \left( \cd_{\mathbf{j}+\mathbf{e}_\mu,\sigma} \c_{\mathbf{j},\sigma} + \hc \right).
\end{equation}
This tunneling perturbation creates excitons, whose $C_4$ angular momentum $m_4=0$ or $2$ can be controlled by choosing $\varphi_y-\varphi_x=0$ or $\pi$, respectively. The perturbation can be written as
\begin{equation}
    \hat{O}_{\mathbf{i}} = \sum_\sigma \sum_{\tau=\pm} \left( \hat{T}_{\tau x}(\mathbf{i},\sigma) + (-1)^{m_4/2} \hat{T}_{\tau y}(\mathbf{i},\sigma) \right),
\end{equation}
for these two values of $m_4=0,2$ and using Eq.~\eqref{eq:DefT} from the main text.

Hence, global lattice intensity modulation allows to measure zero momentum $s$- and $d$-wave exciton spectra.

\subsection{Lattice shaking}
Another conceptually simple modulation involves application of a uniform force $\mathbf{F}(t)$. This can be achieved by modulating the phases $\varphi_\mu(t) = \delta \varphi_\mu \sin(\omega t)$ of the standing optical waves forming the lattice,
\begin{equation}
    V(\mathbf{r},t) = \sum_{\mu=x,y} V_0 \cos^2 \left( \frac{\pi}{a_\mu} r_\mu - \varphi_\mu(t) \right),
\end{equation}
as described by Tokuno and Giamarchi~\cite{Tokuno2011}; equivalently, an overall harmonic confinement potential can be modulated, coupling to the center-of-mass motion of the atomic cloud, as proposed by Wu et al.~\cite{Wu2015} and realized by Anderson et al.~\cite{Anderson2019}.

In both cases, a coupling to the current operator (Eq.~\eqref{eqDefCurrent} in the main text) is generated, leading to the following modulation on the tight-binding level,
\begin{equation}
    \delta \H = \delta \mathcal{A} \cos(\omega t) \sum_{\mathbf{j}} \sum_{\mu=x,y} \delta \varphi_\mu ~\hat{j}^\mu_{\mathbf{j}}.
\end{equation}
Since the current reads
\begin{equation}
    \hat{j}^\mu_{\mathbf{i}} = i \sum_\sigma \hat{T}_{+\mu}(\mathbf{j},\sigma) + \hc
\end{equation}
in terms of the operators $\hat{T}$ defined in the main text Eq.~\eqref{eq:DefT}, the resulting spectral function $A(\omega,\mathbf{k})$ probes $\mathbf{k}=(0,0)$ and involves $m_4=1,3$ angular momenta ($p$-wave). As well known in the literature, lattice shaking provides a direct measurement of the optical conductivity~\cite{Tokuno2011, Wu2015,Anderson2019}.

\subsection{Standard Raman spectroscopy}
In order to create excitations with non-zero momentum $\mathbf{k}$, Raman transitions can be used. This requires a pair of Raman lasers at momenta $\mathbf{k}_1$ and $\mathbf{k}_2$ and frequencies $\omega_1$ and $\omega_2$, which we assume to be incoherent from the beams forming the primary physics lattice. These lasers create an additional potential modulation of the form
\begin{equation}
    \delta V(\mathbf{r},t) = \delta V ~ \cos^2 \left( \frac{1}{2} \left(\mathbf{k} \cdot \mathbf{r} + \Delta \varphi  +  \omega t \right) \right),
\label{eqRamanLattice}
\end{equation}
where $\mathbf{k} = \mathbf{k}_2 - \mathbf{k}_1$ and $\omega = \omega_2 - \omega_1$ are frequency and wavevector of the modulations and $\Delta \varphi $ is an irrelevant phase difference between the two beams. The modulation frequency $\omega$ and wavevector $\mathbf{k}$ of the resulting excitations in the system can thus be controlled through the Raman beams. 

In order to vary $\mathbf{k}$ one can change the spatial angles of the two Raman beams, although this requires good optical access to the experiment. To circumvent such inconveniences, the high-resolution objective of a quantum gas microscope can be used to focus both Raman beams at well-controlled positions $\mathbf{R}_{1,2} \propto \mathbf{k}_{1,2}$ in the Fourier plane of the objective; this creates running waves at the desired momenta in the image plane of the objective, where the optical lattice modulation is thus realized~\cite{Tai2017}. The positions $\mathbf{R}_{1,2}$ can be easily controlled, allowing for full tunability of the Raman lattice modulation $\mathbf{k}$.

The Raman lattice Eq.~\eqref{eqRamanLattice} directly modulates the on-site potentials in the tight-binding Hubbard model,
\begin{equation}
    \delta \H = - \delta g \sum_\mathbf{j} \sum_\sigma \cos \left( \mathbf{k} \cdot \mathbf{j} + \omega t \right) \hat{n}_{\mathbf{j} \sigma}.
\label{eqHgmodulation}
\end{equation}
To describe how this excites the system, we perform a time-dependent gauge transformation
\begin{equation}
    \hat{U} = \exp \left[ i \frac{\delta g}{\omega} \sum_{\mathbf{j} \sigma} \sin \left( \mathbf{k} \cdot \mathbf{r} + \omega t \right) \hat{n}_{\mathbf{j}\sigma} \right],
\end{equation}
which eliminates the on-site modulation Eq.~\eqref{eqHgmodulation} at the cost of introducing a tunneling Hamiltonian with time-dependent phases $\phi_{\ij}(t)$,
\begin{equation}
    \H_t (t) = - t \sum_\ij \sum_{\sigma} \cd_{\mathbf{i}\sigma} \c_{\mathbf{j}\sigma} e^{i \phi_{\ij}(t)} + \hc, 
\end{equation}
where $t$ is the tunneling amplitude in the original physics lattice. The phases are modulated as follows,
\begin{equation}
    \phi_{\ij}(t) = \underbrace{\frac{2 \delta g}{\omega} \sin \left( \frac{1}{2} \mathbf{k} \cdot (\mathbf{j}-\mathbf{i}) \right)}_{=x_{\ij}} \cos \left( \frac{1}{2} \mathbf{k} \cdot (\mathbf{j} + \mathbf{i}) + \omega t \right).
\end{equation}

To calculate the response of the system to these modulated tunneling phases, we use the identity $e^{i z \cos(\phi)} = \sum_{n=-\infty}^\infty i^n \mathcal{J}_n(z) e^{i n \phi}$. Working to lowest non-zero order in $\delta g \ll |\omega|$, as appropriate for the subsequent linear response calculation, one obtains
\begin{multline}
    e^{i \phi_{\ij}(t)} = \mathcal{J}_0(x_\ij) + \\
    +i \mathcal{J}_1(x_\ij) 2 \cos \left( \frac{1}{2} \mathbf{k} \cdot (\mathbf{j} + \mathbf{i}) + \omega t \right) + \mathcal{O}(\delta g^2)
\end{multline}
Since $\mathcal{J}_0(x_\ij) = 1 + \mathcal{O}(\delta g^2)$ the original tunneling Hamiltonian is recovered; using $\mathcal{J}_1(x) = x/2 + \mathcal{O}(x^3)$ and using the operators $\hat{T}$ from Eq.~\eqref{eq:DefT} in the main text the perturbation term finally becomes
\begin{multline}
    \delta \H = - \delta t \sum_{\mathbf{j}\sigma} \sum_{\mu=x,y} \sum_{\tau=\pm} \sin(k_\mu/2) \\
    \times \cos\left( \mathbf{k} \cdot \mathbf{j} + \omega t + \tau k_\mu /2 \right) ~ i \tau ~ \hat{T}_{\tau \mu}(\mathbf{j},\sigma),
\end{multline}
where $\delta t = 2 t \delta g/\omega$.

From the last equation we read off that Raman spectroscopy allows to probe superpositions of the $m_4$-spectra defined in the main text. For momenta around $\mathbf{k} \approx (0,0)$ and $(\pi,\pi)$ modulations along $x$- and $y$-directions are in-phase, whereas for momenta around $\mathbf{k} \approx (0,\pi)$ and $(\pi,0)$ they are $\pi$ out-of-phase. The perturbation always couples to the current operator $\hat{j}^\mu_{\mathbf{j}}$, cf. Eq.~\eqref{eqDefCurrent} in the main text, i.e. Raman spectroscopy only resolves $m_4=0,2$, i.e. p-wave excitations. 

Finally, notice that the factor $\sin(k_\mu/2)$ suppresses any spectral weight around $\mathbf{k}=(0,0)$. In order to measure the optical conductivity at $\mathbf{k}=(0,0)$ global lattice modulation described above must be used.

\subsection{Raman beat spectroscopy}
To address $s$- and $d$-wave excitons at non-zero momenta, modulations coupling directly to the tunneling matrix element instead of the current operators need to be realized. This can be realized by creating modulated optical lattices with nodes at the lattice sites of the primary physics lattice; i.e. these lattices modulate the tunneling barrier between tight-binding Wannier orbitals, coupling directly to the kinetic energy term in the Hubbard Hamiltonian. 

To generate the required optical lattices, one starts with two counter-propagating Raman beams with momentum difference $\mathbf{k}_r^2 - \mathbf{k}_r^1 = \mathbf{K}_x$, where $\mathbf{K}_x$ is the reciprocal lattice vector of the physics lattice along $x$ direction, and equal frequency $\omega_r^1 = \omega_r^2$. By choosing the spatial phase of the resulting standing wave pattern out-of-phase with the primary physics lattice, tunneling strengths along $x$-direction can be modified. Here we assumed no interference with the primary physics lattice, which can be realized by choosing a frequency $\omega_r^1 = \omega_r^2$ sufficiently different from the primary lattice beams. By adding a second pair of beams, non-interfering with the first pair and the primary lattice, hoppings along $y$-direction can be similarly modified.

To realize the slow spatial and temporal modulation of the additional lattice along $\mu$ direction, each Raman pair can be complemented by another pair of coherent Raman beams which are slightly detuned from the first pair by the desired modulation frequency $\omega$. We further assume their momenta $\mathbf{k}_b^1$ and $\mathbf{k}_b^2$ to be given by
\begin{equation}
    \mathbf{k}_b^j = \mathbf{k}_r^j - \mathbf{k}, ~~ j=1,2,
\end{equation}
where $\mathbf{k}$ generates the desired long-wavelength modulation. The latter generates a beat-note of the optical lattice centered at the bonds of the primary physics lattice; for the four beams generating the lattice along $\mu$-direction we obtain a modulated potential
\begin{multline}
    V_\mu(\mathbf{r},t) = \cos^2 \left( \frac{1}{2} \mathbf{K}_\mu \cdot \mathbf{r} \right) ~ \biggl[ 4 (\Omega_r^2 + \Omega_b^2) +\\
    + 8 \Omega_r \Omega_b \cos \left( \mathbf{k} \cdot \mathbf{r} + \omega t + \varphi_\mu \right) \biggr],
\label{eqBeatNoteRaman}
\end{multline}
where $\Omega_r^1=\Omega_r^2=\Omega_r$ and $\Omega_b^1=\Omega_b^2=\Omega_b$ are the Rabi frequencies of the pairs of Raman beams and $\varphi_\mu$ are relative phases of the Raman laser pairs along $x$ and $y$ directions.

The beat note Raman lattice $\sim \Omega_r \Omega_b \propto \delta t$ in Eq.~\eqref{eqBeatNoteRaman} directly modifies the strength of hopping elements in the tight-binding Hamiltonian, i.e.
\begin{multline}
    \delta \H = -\delta t \sum_\mu \sum_{\ij_\mu} \sum_\sigma \cos \left( \frac{1}{2} \mathbf{k} \cdot (\mathbf{i}+\mathbf{j}) + \omega t +\varphi_\mu \right) \\
    \times \cd_{\mathbf{i}\sigma} \c_{\mathbf{j}\sigma} + \hc,
\end{multline}
where $\ij_\mu$ denotes nearest-neighbor bonds along $\mu$-direction. Using the operators $\hat{T}$ from Eq.~\eqref{eq:DefT} in the main text the perturbation becomes
\begin{multline}
    \delta \H = - \delta t \sum_{\mathbf{j} \sigma } \sum_{\mu=x,y} \sum_{\tau=\pm} \cos \left(\mathbf{k} \cdot \mathbf{j} + \omega t + \tau k_\mu /2 +\varphi_\mu \right) \\
    \times \hat{T}_{\tau\mu}(\mathbf{j},\sigma).
\end{multline}
Using the phase difference $\varphi_x-\varphi_y$ couplings to $s$-wave or $d$-wave excitations can be achieved. The momentum and frequency of the created excitons are freely tunable, without any suppression of matrix elements at specific momenta.

Finally, we note that the beat note Raman spectroscopy can be easily modified to measure $p$-wave excitations in the same setup. To this end, the spatial phase of the standing wave pattern can be choosen to be in-phase with the primary physics lattice, which leads to a modulation of on-site potentials. As described in the previous subsection, the latter generates couplings to $p$-wave excitons.

\section{Details of MPS simulations}
\label{sec:MPSconvergence}

\begin{figure}
\centering
\includegraphics[width=0.95\linewidth]{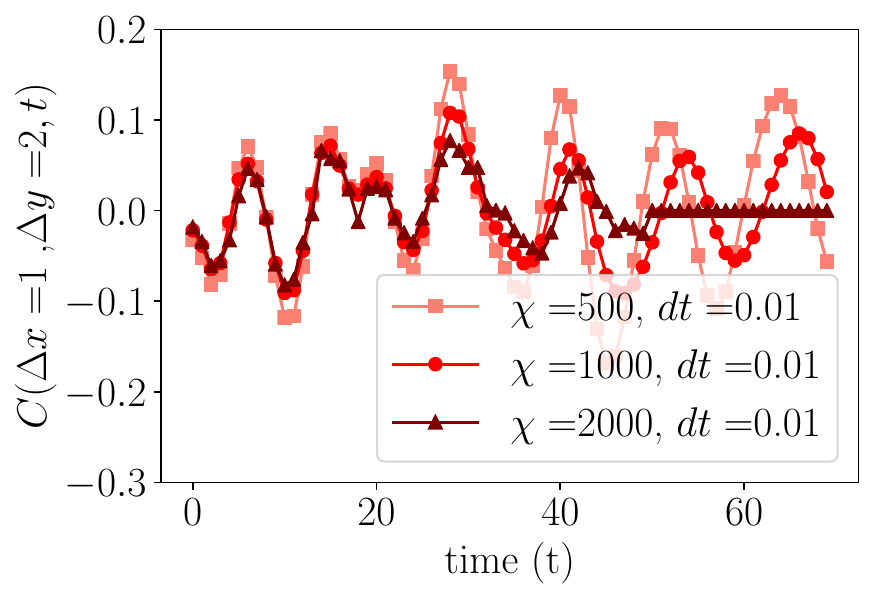} 
\caption{
\textbf{Green's function} in real space and time in the Hubbard model for $U/t=8$, $m_4=2$ at position $(x,y) = (19,2)$ for different bond dimensions $\chi$ and time step $dt=0.01$. Convergence with time step was confirmed independently at fixed $\chi$. 
}
\label{fig:Greensfunction}
\end{figure}

In order to calculate the excitonic and two-dopant spectra shown in the main text, we start from the ground state of the Hubbard / $t-J$ model on a four-leg cylinder (periodic boundary conditions in $y$-direction) at half-filling, and apply the excitation operator Eq.~\ref{eq:DefR}. Subsequently, we time-evolve the matrix product state under the time-independent Hamiltonian using the $W^{II}$ method introduced in Ref.~\cite{Zaletel2015} with a time step of $dt=0.01/t$ and bond dimensions up to $\chi=2000$. We proceed by Fourier transforming to momentum space. We then perform linear extrapolation in time, and multiply the resulting signal with a Gaussian function, see also Refs.~\cite{Bohrdt2020_ARPES,Bohrdt2023}.
We ensured convergence in the bond dimension and time step, as shown exemplary for the Green's function Eq.~\ref{eq:rotationalGF} in real space and time for the Hubbard model at $U/t=8$, $m_4=2$ in Fig.~\ref{fig:Greensfunction} and use times up to $T=4.0/t$.

\section{Excitonic spectra at different $k_y$ and $m_4$ } \label{sec:MoreSpectra}

\begin{figure*}
\centering
\includegraphics[width=0.9\linewidth]{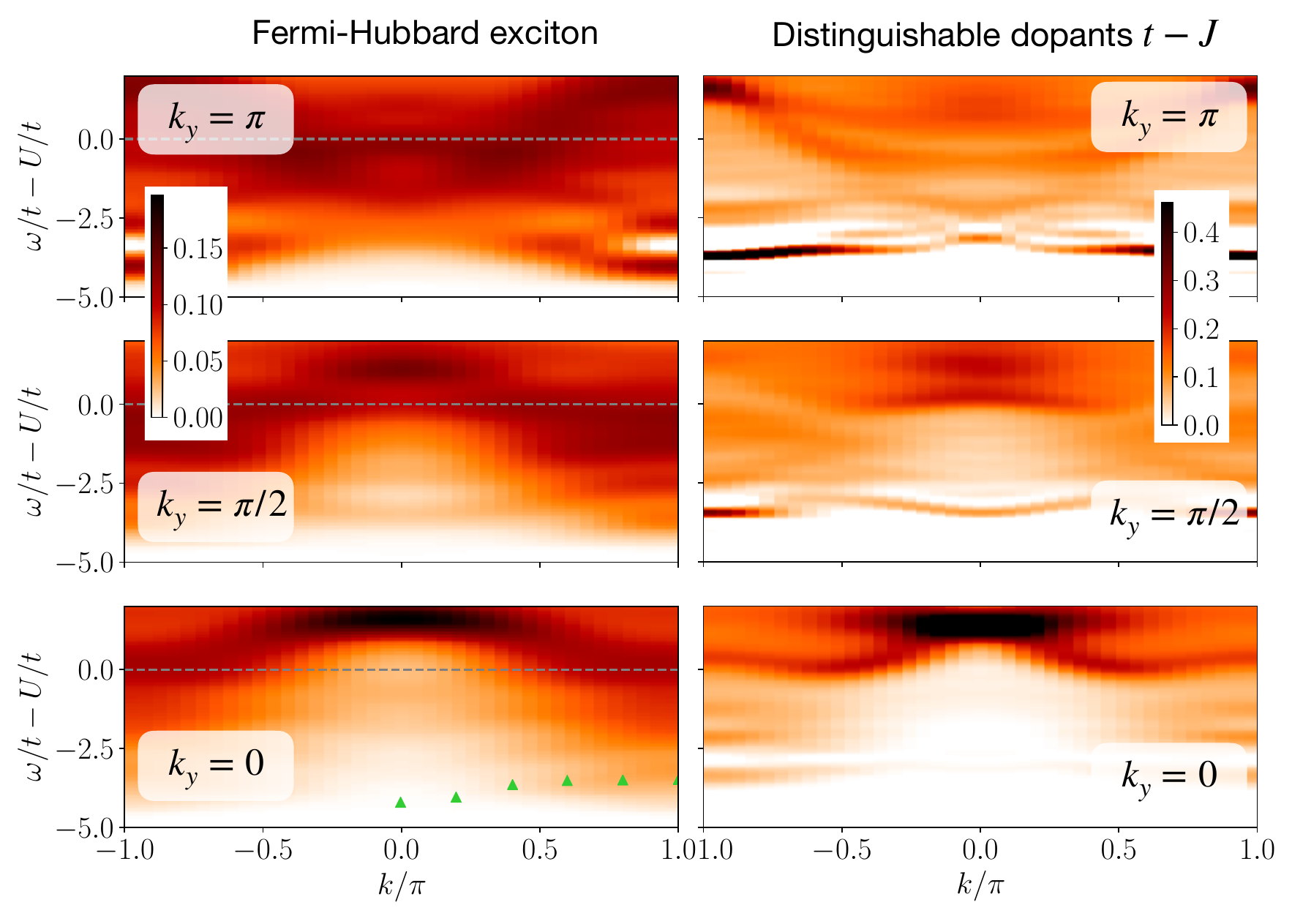} 
\caption{
\textbf{$s$-wave exciton spectra} with momentum resolution in $x$-direction at $k_y=0$ (bottom), $k_y=\pi/2$ (middle), and $k_y=\pi$ (top) in the Fermi-Hubbard model for $U/t=12$ (left) and two hole spectra in the $t-J$ model with distinguishable dopants for $t/J=3$ (right) for an $m_4=0$ excitation.  In the Fermi-Hubbard model for $k_y=0$, we show the peak positions as extracted from RIXS measurements on $\text{La}_2\text{CuO}_4$ \cite{Ellis2008} (triangles), assuming $t=350\text{meV}$. 
}
\label{fig:spectrum_dist_m0}
\end{figure*}

\begin{figure*}
\centering
\includegraphics[width=0.9\linewidth]{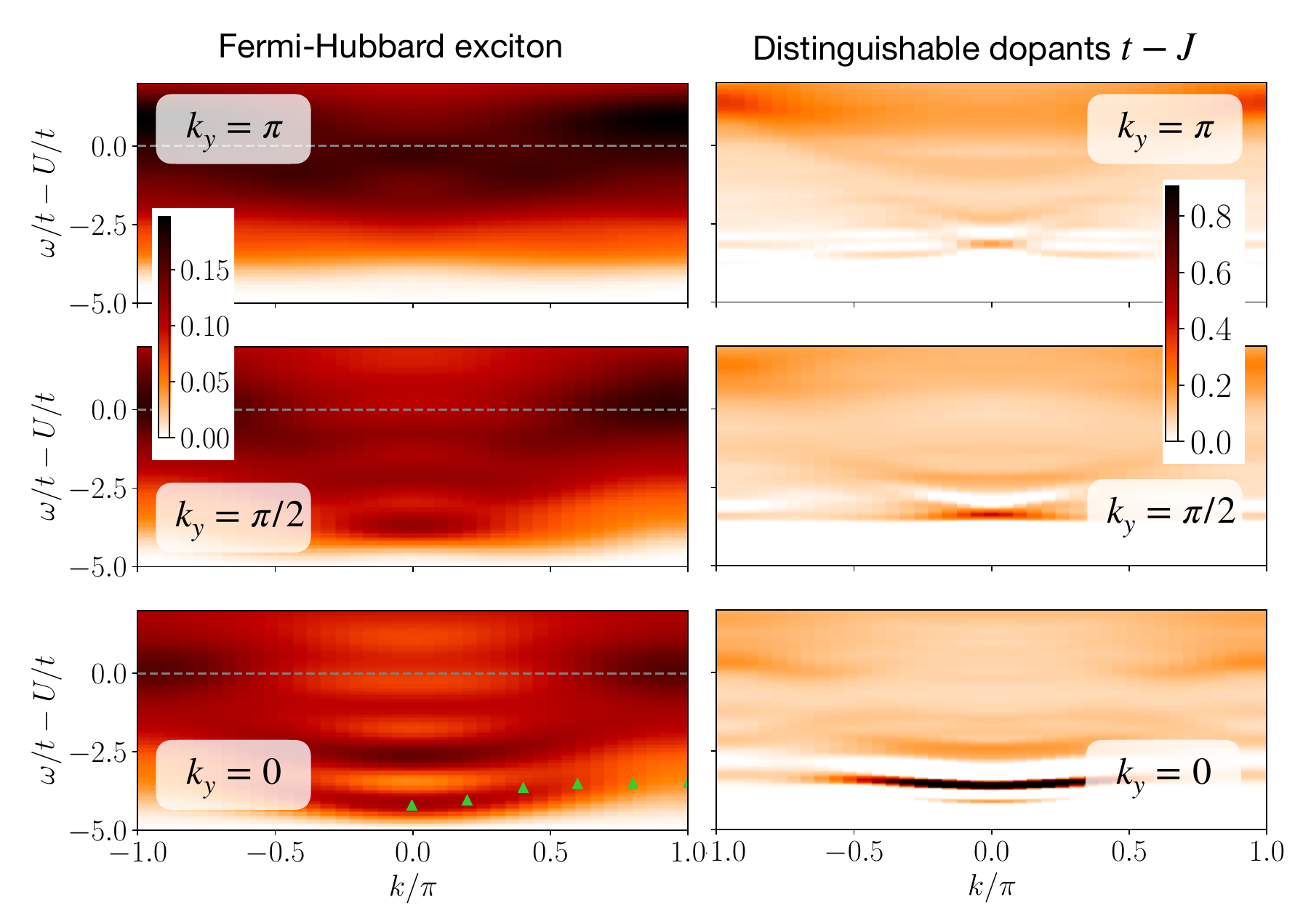} 
\caption{
\textbf{$d$-wave exciton spectra} with momentum resolution in $x$-direction at $k_y=0$ (bottom), $k_y=\pi/2$ (middle), and $k_y=\pi$ (top) in the Fermi-Hubbard model for $U/t=12$ (left) and two hole spectra in the $t-J$ model with distinguishable dopants for $t/J=3$ (right) for an $m_4=2$ excitation.  In the Fermi-Hubbard model for $k_y=0$, we show the peak positions as extracted from RIXS measurements on $\text{La}_2\text{CuO}_4$ \cite{Ellis2008} (triangles), assuming $t=350\text{meV}$. 
}
\label{fig:spectrum_dist_m2}
\end{figure*}

In Figs.~\ref{fig:spectrum_dist_m0} and \ref{fig:spectrum_dist_m2}, we show the full numerical spectra with all available momentum points for the exciton in the Hubbard model and two distinguishable dopants in the $t-J$ model for $m_4=0$ and $m_4=2$, respectively. We observe similar features in the Hubbard and $t-J$ model, such as e.g. an avoided crossing at low energies for $m_4=0$ around $\vec{k}=(\pi/2,\pi/2)$, a suppression of spectral weight for $m_4=0$ and $\vec{k}=(0,0)$, and various branches e.g. at $\vec{k}=(\pi,\pi)$ and $(0,\pi)$.


\end{document}